\documentclass{template/svproc}
\usepackage{url}
\usepackage{graphicx}
\usepackage{subcaption}
\usepackage{cite}
\usepackage{amsmath}
\usepackage{xcolor}
\usepackage{adjustbox} % Para ajustar el tamaño de las figuras y moverlas
\usepackage{tikz}  % Paquete para dibujar figuras

\begin{document}
\mainmatter              % start of a contribution
\title{A Data-Centric Framework for Machine Listening Projects: Addressing Large-Scale Data Acquisition and Labeling through Active Learning}
\titlerunning{A Data-Centric Framework for Machine Listening Projects}  % abbreviated title (for running head)
\author{Javier Naranjo-Alcazar \and Jordi Grau-Haro \and
Ruben Ribes-Serrano \and Pedro Zuccarello}
\authorrunning{Javier Naranjo-Alcazar et al.} % abbreviated author list (for running head)
\tocauthor{Javier Naranjo-Alcazar, Jordi Grau-Haro, Ruben Ribes-Serrano and Pedro Zuccarello}
\institute{Instituto Tecnológico de Informática (ITI), Valencia, Spain \\
\email{\{jnaranjo, jgrau, rribes, pzuccarello\}@iti.es}
}
% \and
% Universit\'{e} de Paris-Sud,
% Laboratoire d'Analyse Num\'{e}rique, B\^{a}timent 425,\\
% F-91405 Orsay Cedex, France}

\maketitle              % typeset the title of the contribution

\begin{abstract}
Machine Listening focuses on developing technologies to extract relevant information from audio signals. A critical aspect of these projects is the acquisition and labeling of contextualized data, which is inherently complex and requires specific resources and strategies. Despite the availability of some audio datasets, many are unsuitable for commercial applications. The paper emphasizes the importance of Active Learning (AL) using expert labelers over crowdsourcing, which often lacks detailed insights into dataset structures. AL is an iterative process combining human labelers and AI models to optimize the labeling budget by intelligently selecting samples for human review. This approach addresses the challenge of handling large, constantly growing datasets that exceed available computational resources and memory. The paper presents a comprehensive data-centric framework for Machine Listening projects, detailing the configuration of recording nodes, database structure, and labeling budget optimization in resource-constrained scenarios. Applied to an industrial port in Valencia, Spain, the framework successfully labeled 6540 ten-second audio samples over five months with a small team, demonstrating its effectiveness and adaptability to various resource availability situations.
% We would like to encourage you to list your keywords within
% the abstract section using the \keywords{...} command.
\keywords{Machine Listening, Active Learning, Artificial Intelligence, Dataset, IoT}
\end{abstract}
\section{Introduction}\label{sec:intro}

Machine Listening is the field of knowledge that aims to create technological solutions that are able to obtain relevant information from audio signals. Some of the most common problems present in this field can be  Acoustic Scene Classification \cite{Schmid2023}, Sound Event Detection \cite{mesaros2021sound, kandewal} or Anomalous Sound Detection \cite{wilkinghoff2024self} among others \cite{Aguado2023}. 
%kandewal in sound event detection

% se puede añadir
%AAC can be defined as the problem that aims at creating a description of an audio clip through a sentence. ASC corresponds to the classification of an audio clip into a pre-defined scene. SED solutions can find sound events in an audio clip indicating their start and end times. ASD solutions are intended to predict anomalies in the test phase, being that only sounds considered normal were used in the training phase.

One of the main pillars of any Machine Listening project, and of any other AI project, is the acquisition and labeling of relevant and contextualized data for the problem to be addressed. This task is usually not simple and needs specific resources and strategies to be carried out successfully. There are several projects in the literature related to the creation of generic sound databases to be used in Machine Listening projects. The Sounds Of New York (SONYC) \cite{mydlarz2019life} project is one of them, in which about 50 recording nodes were deployed throughout various areas of New York City. Other similar projects, in which nodes have also been deployed in areas of high urban concentration, are those of Montevideo \cite{Zinemanas2019} and Singapore \cite{ooi2021strongly}.

Although there are some audio datasets, they are not usually suitable for commercial purposes. The most widely used in the scientific literature is Audioset \cite{gemmeke2017audio} which is composed of audios from YouTube videos. This dataset, while extensive, does not give access to the audios themselves but only to the feature vectors. Furthermore, it can only be used for research activities, never for commercial uses and purposes. Even though some of them can be used for commercial purposes, no updates and/or improvements to these datasets have been released for a long time \cite{piczak2015dataset, salamon2014dataset}.

As for labeling, two common strategies are observed: Active Learning (AL) based on expert opinions or crowdsourcing \cite{martin2023strong, lipping2019crowdsourcing, martin2021crowdsourcing}. In the case of crowdsourcing, the usual scenario is that the labelers are not experts on the topic, so that no real insight into the structure and details of the dataset is obtained. In the opinion of the authors of this paper, the best option is to work through AL iterations with a pool of expert human labelers.

Since the usual scenario is that the amount of data to be labeled far exceeds the resources available to perform the task, the procedure known as Active Learning aims at optimizing the labeling budget. Active Learning is an iterative procedure that combines human labelers and AI models in such a way that these models smartly select which samples should be listened and labeled by humans. This selection is done by optimizing a certain criterion based on the principles of uncertainty, discrepancy of a model committee, estimation of change in the model, among others \cite{settles2009active}. In published scientific work on AL for audio, it is common for the dataset to have a controlled and computationally manageable size, both from the point of view of computational complexity and memory space. In our case study, this is not the case because the dataset grows constantly so that there comes a point when both the computational resources and the memory available are not enough to cope with the necessary calculations on the entire dataset. In this paper a series of measures are proposed and implemented in order to deal with this situation.

% Dibujar el rectángulo con TikZ
\begin{tikzpicture}[remember picture, overlay]
    \node at (current page.south) [anchor=south, yshift=1.5cm] {\textbf{Paper accepted at 8th Future of Information and Communication Conference 2025, 28-29 April, Berlin}};
\end{tikzpicture}

\begin{figure}[]
    \centering
    \begin{subfigure}[b]{0.45\textwidth}
        \centering
        \adjustbox{raise=1cm}{ % Ajusta el valor para elevar la imagen y el caption
        \includegraphics[width=\textwidth]{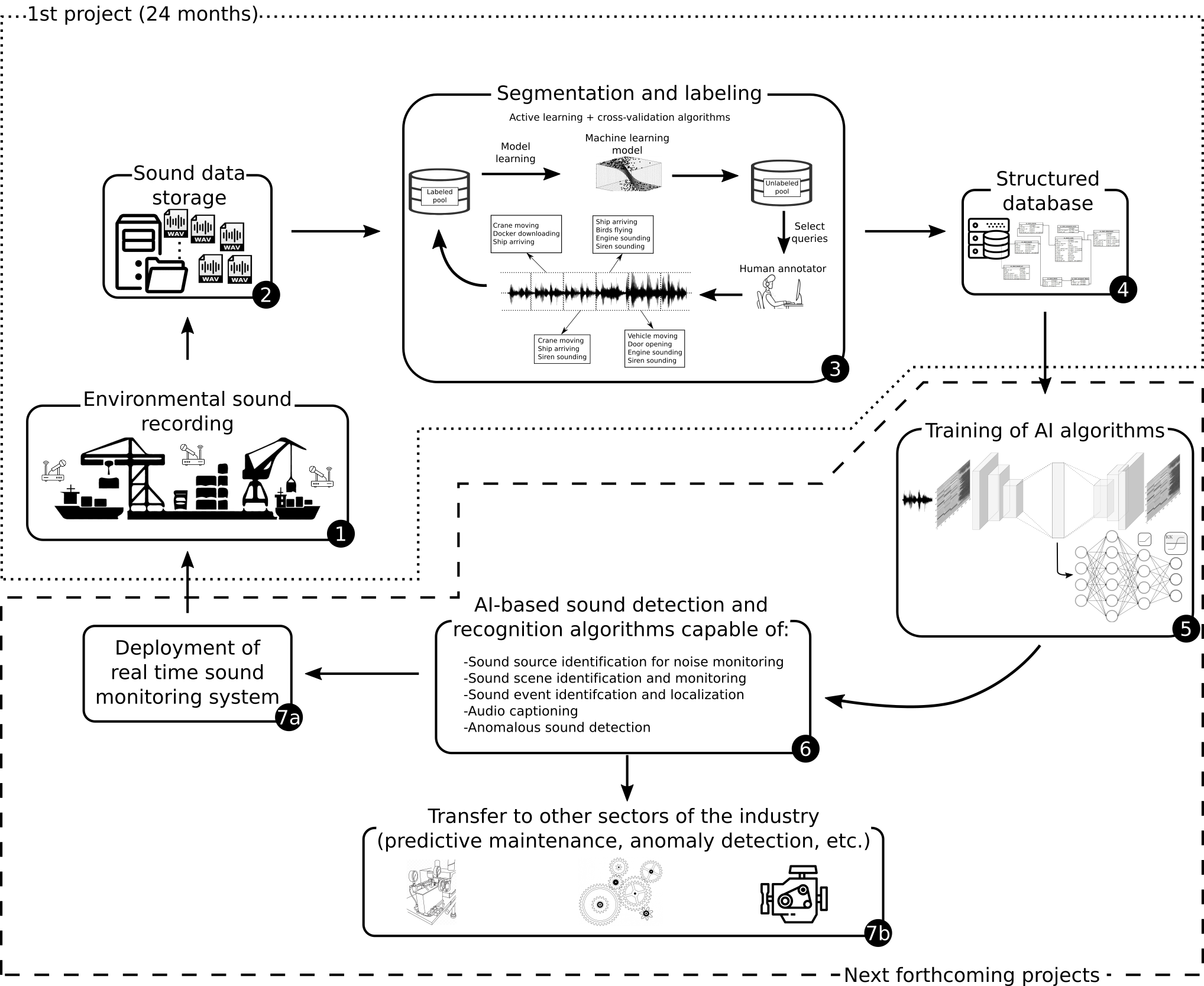}}
        \caption{}
        \label{fig:planning}
    \end{subfigure}
    \hfill
    \begin{subfigure}[b]{0.45\textwidth}
        \centering
        \begin{subfigure}[b]{\textwidth}
            \centering
            \includegraphics[width=0.75\textwidth]{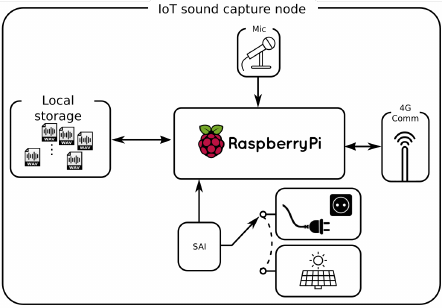}
            \caption{}
            \label{fig:suba}
        \end{subfigure}
        \begin{subfigure}[b]{\textwidth}
            \centering
            \includegraphics[width=0.75\textwidth]{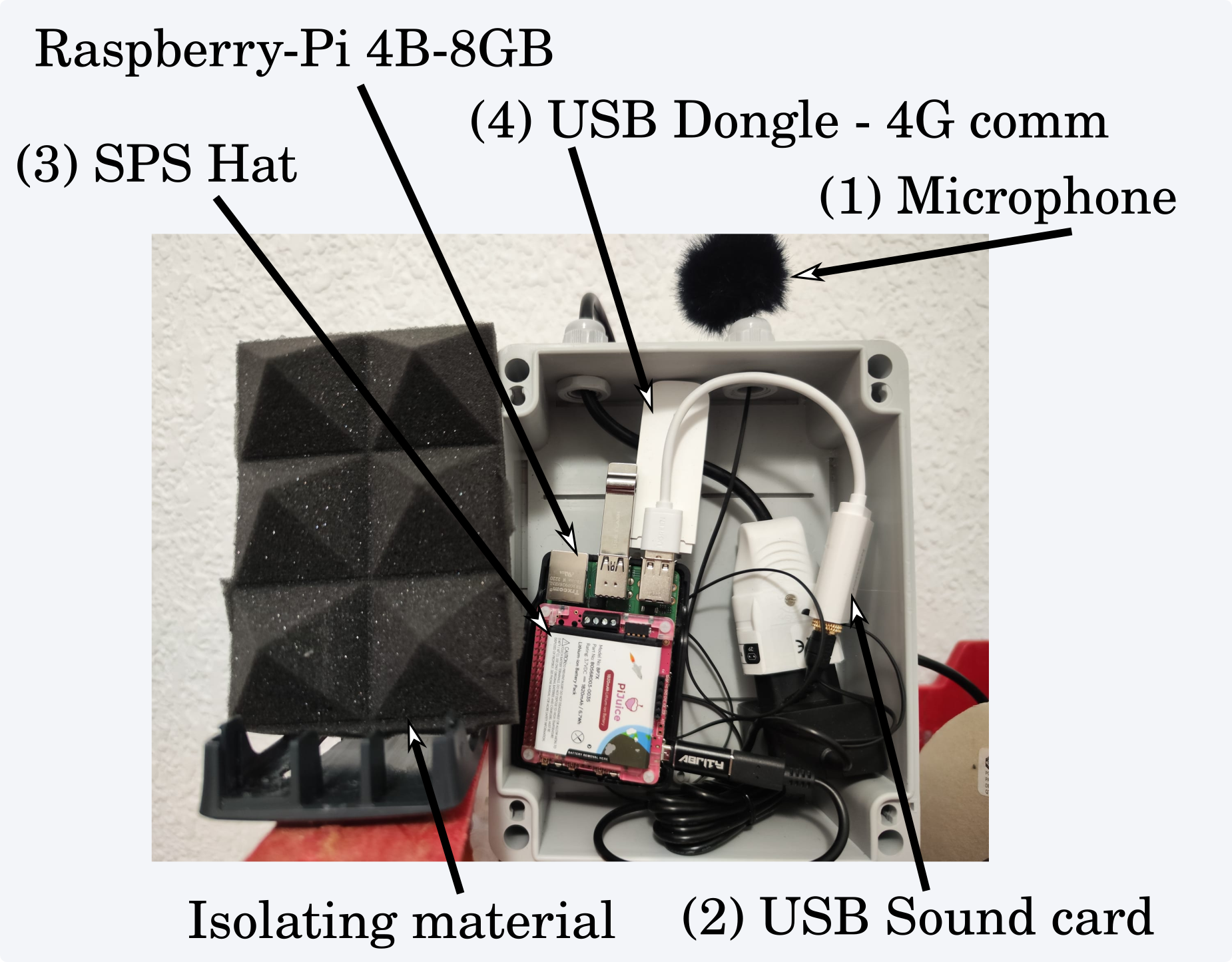}
            \caption{}
            \label{fig:subb}
        \end{subfigure}
        %\caption{Recording node desigh}
        %\label{fig:subb}
    \end{subfigure}
    \caption{(a) Steps to be taken to solve Machine Listening problems with a Data-Centric perspective. (b) Recording node block diagram. (c) Recording node real ensemble}
    \label{fig:al_node}
\end{figure}

This paper presents a comprehensive framework for approaching Machine Listening projects from a Data-Centric perspective (see Figure~\ref{fig:planning}). The work addresses the problems that arise when trying to translate AL algorithms and procedures from state-of-the-art papers to a real case of large-quantity continuous sound-data collection and labeling projects. The project on which this paper is based has been ongoing for the last year and a half and is expected to last 3 years in total. The deployment reaches between 3 and 7 IoT nodes that capture sound 24 hours a day. When the amount of data is so large, different problems arise when trying to directly apply the AL algorithms present in the literature. The paper describes the problems encountered and explains the solutions adopted to solve them. 

This work presents the configuration of the recording node, the structure of the database that compose the dataset and the optimization of the labeling budget in a scenario of constrained resources (small labelers team) when it is intended to avoid crowdsourcing. The presented framework can be parameterized to be modified depending on the particular case and availability of resources. The results shown in this work are obtained from the use case applied to the context of an industrial commercial port located in the city of Valencia (Spain) where 7 recording nodes are currently deployed. At the time of writing this paper, 6540 audios of 10 seconds duration have been labeled during 5 months (since January 8, 2024) with an effort of 30 minutes of labeling per day, which corresponds to each tagger labeling 40 audios per day. The human labelers teams is composed of five people which have been divided into two groups in order to perform cross-validation techniques and avoid biases.

The paper does not propose a new AL algorithm. Rather, it shows a complete framework for the creation of a labeled audio database in a context where the labeling budget is clearly lower than the amount of data collected. Moreover, it proposes how to deal with the limitations in computational capacity related to the amount of data to be processed by the AL algorithm. The whole proposal of the paper is an essential preliminary step to the training of future AI models in any machine listening project. Working with an unlabeled dataset, i.e., without a labeled ground-truth, makes comparison with other AL algorithms impossible. Therefore, the paper focuses on the necessary practical aspects of this kind of projects. 

The paper is structured as follows. Section 2 is divided in several subsections in which the following items are explained: \ref{subsec:node} the recording node, \ref{subsec:al_intro} the Active Learning algorithm used in this work as well as an overview of the state of the art, \ref{subsec:database} the database architecture used to store the recording and labeling process information and finally \ref{subsec:al} the framework proposed in this work. Section~\ref{sec:results} shows the results obtained after several iterations of the proposed pipeline showing its performance. Finally, Section~\ref{sec:conclusions} concludes our work and introduces future steps in the research line.

\section{Materials and Methods}\label{sec:method}

The framework is formed by: the recording node, the raw storage of the audio data, the AL process and the building of a structured database with the labeled data. The following sections present each of the modules.

\subsection{Recording node}\label{subsec:node}

The core of the IoT device for recording is a Raspberry Pi card. Additional peripheral devices were added to improve remote monitoring and avoid possible failures in the event of a power outage (see Figures~\ref{fig:suba} and \ref{fig:subb}): (i) a Rode Lavalier II microphone, (ii) a USB sound card to connect the microphone to a USB slot, (iii) an SPS Hat so that a safe shutdown occurs in the event of power line supply failure and (iv) a USB router with internet connectivity through a SIM card to remotely monitor the node's activity. In the case of power supply through solar panels, the SPS Hat is indispensable as it allows setting hardware on and off times. Once assembled, the device is placed inside a PVC IP65 box with two openings: one for the microphone and the other for the power supply cable.

When recording, files are saved every 10 seconds. The timestamp of the 10-second chunk is implicit in its filename. Remote monitoring of the nodes was implemented via the periodic transmission of different node state variables using the MQTT protocol. A ThingsBoard panel was implemented to visualize these states. To detect possible failures of the microphone, a periodic sending of an audio file every hour has been scheduled  to a Minio bucket. Thus, every day, a specific number of audios from each node are listened to verify the microphone status. For illustrative purposes, a repository with the described functionalities has been released\footnote{\url{https://github.com/JNaranjo-Alcazar/recording_node_monitoring.git}}.

\subsection{Active Learning}\label{subsec:al_intro}

Several AL solutions have been proposed in recent years in the context of Machine Listening solutions.  The line of works that concludes with \cite{9217930} can be understood as an evolution of an AL algorithm intended for audio classification. In \cite{7952256} an AL method called Medoid Based AL (MAL) was proposed based on performing an unsupervised K-medoid clustering on unlabeled data. The medoids are then proposed to be labeled by human annotators. In each iteration, the medoids are excluded from the next proposal. The next work, called MAL-MF (Mismatch-First) \cite{8521336}, can be considered an improvement of the MAL method. In MAL-MF, MAL is used for the first iteration, when there is no human-labeled data. Once labeled data is available, a classifier is trained. From the second iteration onwards, a committee-based approach is used where samples with discrepancies between the classifier and the medoids propagation are candidates for labeling. Samples with discrepancies are sorted according to a distance criterion for diversity consideration.

Other approaches are based on the uncertainty of a certain classifier \cite{settles2009active}. The main idea of uncertainty-based approaches is that the unlabeled samples with the lowest classification probability to a specific class should be labeled next. The classifier is trained iteratively as new labeled samples are available. Works that employ this method include using Monte Carlo dropout \cite{shishkin2021active}, Gaussian distribution for estimated informativeness \cite{shishkin2024active}  or \cite{Meire2023} which studies, in addition to uncertainty, a hybrid approach using semi-supervised detection of outliers techniques. Although uncertainty-based approximations provide very informative samples, these approximations can lead to bias due to mislabeling. To avoid this phenomenon, in \cite{8683063} a solution is proposed to label certain samples every a specified number of iterations.

In this work the MAL-MF was implemented. However, uncertainty and diversity principles  with a pre-trained network are used to optimize the AL procedure when a large amount of data is available. This is known as batch AL \cite{batch}. Although, the principles of uncertainty and diversity have shown satisfactory results in computer vision. To the best of this group's knowledge, no batch AL approach has been performed in the field of machine listening using a pre-trained network as a basis for the above mentioned principles. Information from each AL iteration is stored in an intermediate database. 

%\textcolor{blue}{Esto es conocido como batch AL (CITA). Si bien, los principios de incertiumbre  y diversidad than demostrado resultados satisfactorios en computer vision (CITA). Hasta donde llega el conocimiento de este grupo, no se ha realizado una aproximación de batch AL en campo del machine listening utilizando una red pre-entrenada como base para los principios previamente mencionados.}

\subsubsection{Active Learning Database}\label{subsubsec:al_database}

The tables that compose this database are:

\begin{itemize}
        \item \textit{ALPreprocessing}: information about the time window for analysis, the number of audios, the number of folds, the date and the percentage of labeled data that contains the execution (see step 1 in Fig.\ref{fig:al_framework}).
        
        \item \textit{WavsProposed}: information relevant to the audios proposed by the AL algorithm. It contains two foreign keys: (i) the ID of the AL iteration (id of \textit{ALPreprocessing}) and (ii) the id of the proposed audio from the table \textit{Audios}. Other fields, such as the label, the number of labelers who have tagged the audio and the percentage of agreement of the labelers to tag the audio, the file name and the node's id are also included.
\end{itemize}

\subsection{Audio Database}\label{subsec:database}

\begin{figure}
    \centering
    \includegraphics[width=1\linewidth]{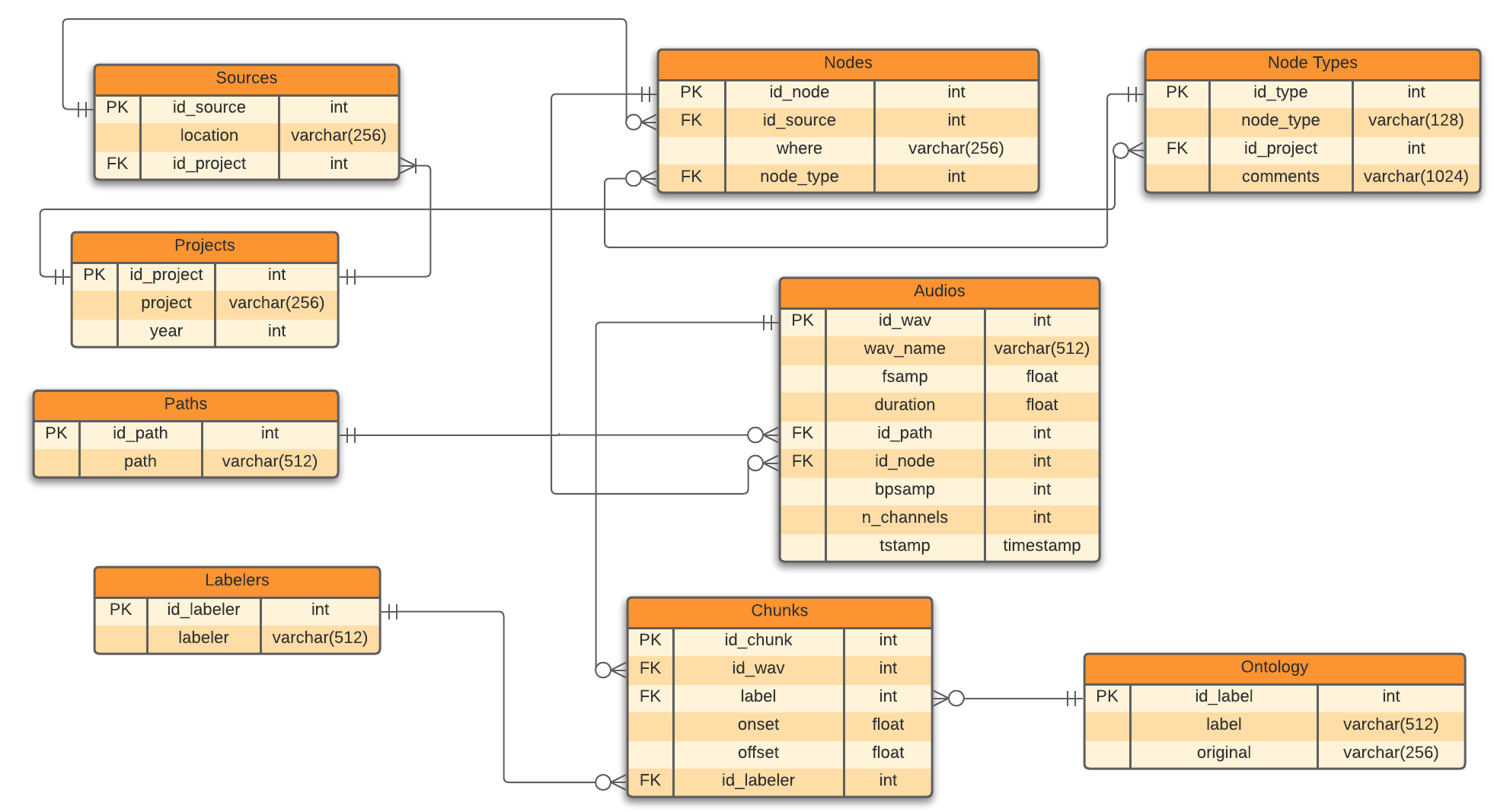}
    \caption{Audio Database structure}
    \label{fig:database}
\end{figure}

This database is composed of the information obtained in the labeling process (see Fig~\ref{fig:database}). It is constituted by the following tables:

\begin{itemize}
        \item \textit{Projects}: list of the projects in which audios have been captured to be incorporated into the database
        \item \textit{Sources}: this table refers to the physical locations where the recording nodes are deployed
        \item \textit{Node Types}: information about the different types of deployed recording nodes. In our case two different registers refer to the two ways of powering the nodes: by solar panels or by mains electricity.
        \item \textit{Nodes}: information associated to each deployed IoT Recording Node. Each node has two foreign keys referring to the tables \textit{Sources} and \emph{Node Types}.
        \item \textit{Paths}: paths of the folders where the audios are stored.
        \item \textit{Ontology}: tags present in the database
        \item \textit{Labelers}: Id of the human labelers 
        \item \textit{Audios}: information about the audios captured by the recording nodes such us filename, sampling rate, bits per sample, duration, or number of channels. It has two foreign keys: (i) the path's and (ii) node's id.
        \item \textit{Chunks}: all audio events identified in the labeling process. It contains three foreign keys: (i) id of the corresponding audio file, (ii) id of the event label from the Ontology table and (iii) the labeler's id. In the case of labeling the beginning and end of the event (strong labeling), two fields corresponding to the onset/offset times of the event must be added.
    \end{itemize}

\subsection{Detailed Active Learning Framework}\label{subsec:al}

\begin{figure}
    \centering
    \includegraphics[width=1\linewidth]{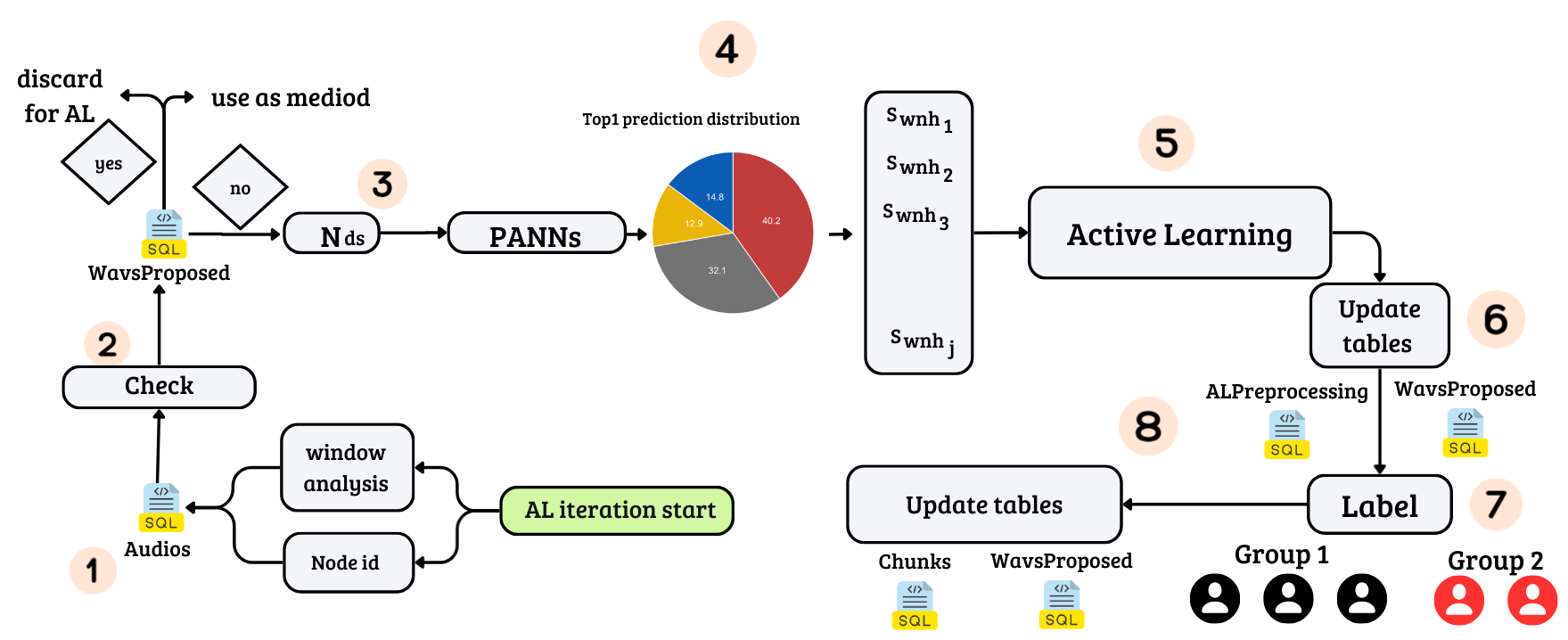}
    \caption{Framework for creating an audio database with a large amount of available data while optimizing the labeling budget}
    \label{fig:al_framework}
\end{figure}

Our AL implementation follows the one proposed in \cite{8521336}. Many of the AL algorithms in the literature are suitable only for problems with small-scale datasets where all the samples can be used for calculations in each iteration of the AL procedure. However, the presented scenario faces a huge and constantly growing unlabeled dataset, thus, there will come a point where the algorithm cannot be applied to the entire data set due to computational capacity limitations. Therefore, measures must be taken to deal with this situation. One possibility is to perform the AL iterations over time windows, e.g., a particular month, a particular week, etc. (step 1 in Figure~\ref{fig:al_framework}). Even with this restriction the amount of data can still be very large. As a solution to this, the procedure proposed, and implemented, in this paper is based on dividing the samples into different disjoint sets. The criterion to assemble the sets is based on the variability shown in the CNN14 PANNs top-1 predictions (the class with the highest probability), where PANNs is a collection of models of different CNNs with public checkpoints trained with Audioset \cite{kong2020panns}. The sets are located in different disjoint sets in such a way that the first disjoint set is more informative and diverse than the last one.  This division process is depicted in steps 3 and 4 of Figure~\ref{fig:al_framework} and is explained in more detail below. The available computational capacity constrains the quantity of samples in each disjoint set.

The first classifier of the committee corresponds to the propagation of labels from human-labeled samples, which are considered the medoids of a given cluster, to their nearest unlabeled neighbors. The second classifier is a Logistic Regression. The feature vector used to train both, the Logistic Regression and the medoid propagation, is the PANNs pre-trained network CNN14 embedding. An uncertainty approach is used for the division of samples among disjoint sets. The steps performed for an AL iteration  (see Figure~\ref{fig:al_framework}) are:

\begin{enumerate}
    \item Due to the large amount of data available, at the beginning of each AL iteration, a selection of the samples to be processed is made. This filter is performed by setting a time window on the set of data to be processed. In our case, we assume that the data of interest are those belonging to a particular recording node. We call $S_w$ to the set of samples selected in this way.
   
    \item The $S_w$ set is divided into two subgroups depending on the existence of previous labeling over the same analysis window in previous AL iterations. Samples that have been labeled form the medoid group, $S_{wm}$. On the other hand, we call $S_{wnh}$ to the group of samples that belong to $S_w$ and that have not been labeled. It is verified that $S_w = S_{wm} \cup S_{wnh}$ and $S_{wm}\cap S_{wnh} = \emptyset$.
    
    \item Due to limitations in the computational capacity of the available hardware, the number of samples in $S_{wnh}$ may be larger than the number of samples that can be processed by the AL algorithm. Thus, a partitioning of $S_{wnh}$ into several disjoint sets is performed. We denote $N_{Smax}$ as the maximum number of samples that can be processed by the AL algorithm. In our implementation, $N_{Smax}=15000$. Therefore, the number of disjoint sets, $N_{ds}$, is calculated as $N_{ds}=\lvert S_{wnh}\rvert / N_{Smax}$. Let $S_{wnh-j}$ be the $j$-th disjoint set, with $1\leq j\leq N_{ds}$. The procedure followed to assign the samples to each set $S_{wnh-j}$ is as follows:
    \begin{enumerate}
        \item The PANNs network is used to assign a first label/classification to the $S_{wnh}$ samples. This network can detect multiple events in the audio segment and therefore assign multiple labels. However, to perform the separation into disjoint sets, it is considered that only a single label is needed. It has been decided that this is the one that PANNs points to as having the highest probability, this is, the PANNs top-1 prediction. Let $L_c$ be the number of samples to which PANNs has assigned class/label $c$ as top-1.
    
        \item The different disjoint sets will be formed following a decreasing principle of uncertainty and diversity in such a way that if $P_c(x) < P_c(y)$ and $x\in S_{wnh-i}$ and $y\in S_{wnh-j}$, then $i<j$, where $P_c(x)$ and $P_c(y)$ are, respectively, the probabilities that samples $x$ and $y$ belong to class $c$ according to PANNs top-1 prediction. Thus, the priority sets will be those with the highest uncertainty and diversity. To assign the samples of a given PANNs top-1 class, $c$, the following three rules have been followed:
        \begin{enumerate}
            \item If $L_c < N_{ds}$, then all the samples of that class are added to the set $S_{wnh-1}$, this being the set with the highest priority.
            
            \item If $L_c = N_{ds}$, one sample is added to each group $S_{wnh-j}$, with  $1 \leq j \leq N_{ds}$. Samples with lower probability, i.e. higher uncertainty, are added to the higher priority sets, i.e. with lower subscript.
            
            \item If $L_c > N_{ds}$, then, $L_c/N_{ds}$ samples are added to each set following the same uncertainty principle described for the case of $L_c=N_{ds}$.
        \end{enumerate}
    \end{enumerate}

    \item The number of medoids can also be a limiting factor for the computational capacity. Thus, a selection criterion must be used. In our implementation, the priority medodis are the ones in set $S_{wm}$. In case more medoids can be added, labeled samples from the same node but from different time windows are added. Finally, samples from other nodes already labeled would be added. After the selection of medoids the AL algorithm is launched on the selected disjoint set \cite{8521336}.
    \item AL database tables are updated
    \item The samples proposed by the AL algorithm are provided to the human-labeler teams. Our own modified version of BAT\footnote{\url{https://github.com/BlaiMelendezCatalan/BAT}} has been used as the labeling tool in this work.
    \item \textit{Chunks} table of the database is updated as well as the medoids field of the \textit{ProposedWavs} table of the AL database. Since several tags may appear in a single audio, a strategy must be defined to define which is the medoid class of the audio for propagation in the next iteration of AL. The criterion followed for updating the medoids field with a label depends on the agreement between the labelers within its group. This agreement must satisfy two requirements: (i) the labels have to be in, at least, 2/3 of the labelers' annotations within the group and (ii) if more than one label satisfies this requirement, the selected label will be the one that lasts the longest.
\end{enumerate}

\section{Result}\label{sec:results}

\begin{figure}
    \centering
    \begin{subfigure}[b]{0.49\textwidth}
        \centering
        \includegraphics[width=\textwidth]{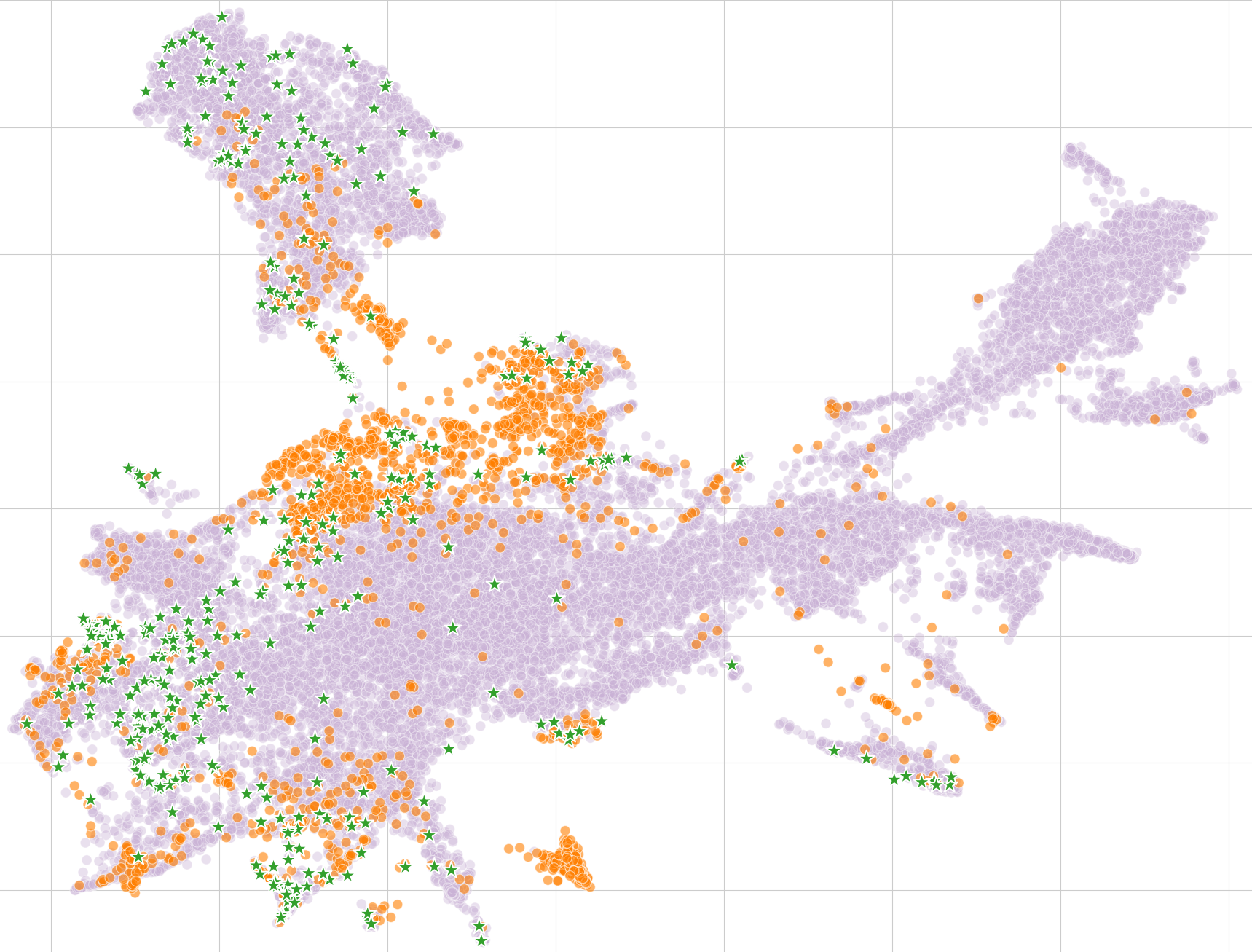}
        \caption{}
        \label{fig:suba_umap}
    \end{subfigure}
    \hfill
    \begin{subfigure}[b]{0.49\textwidth}
        \centering
        \includegraphics[width=\textwidth]{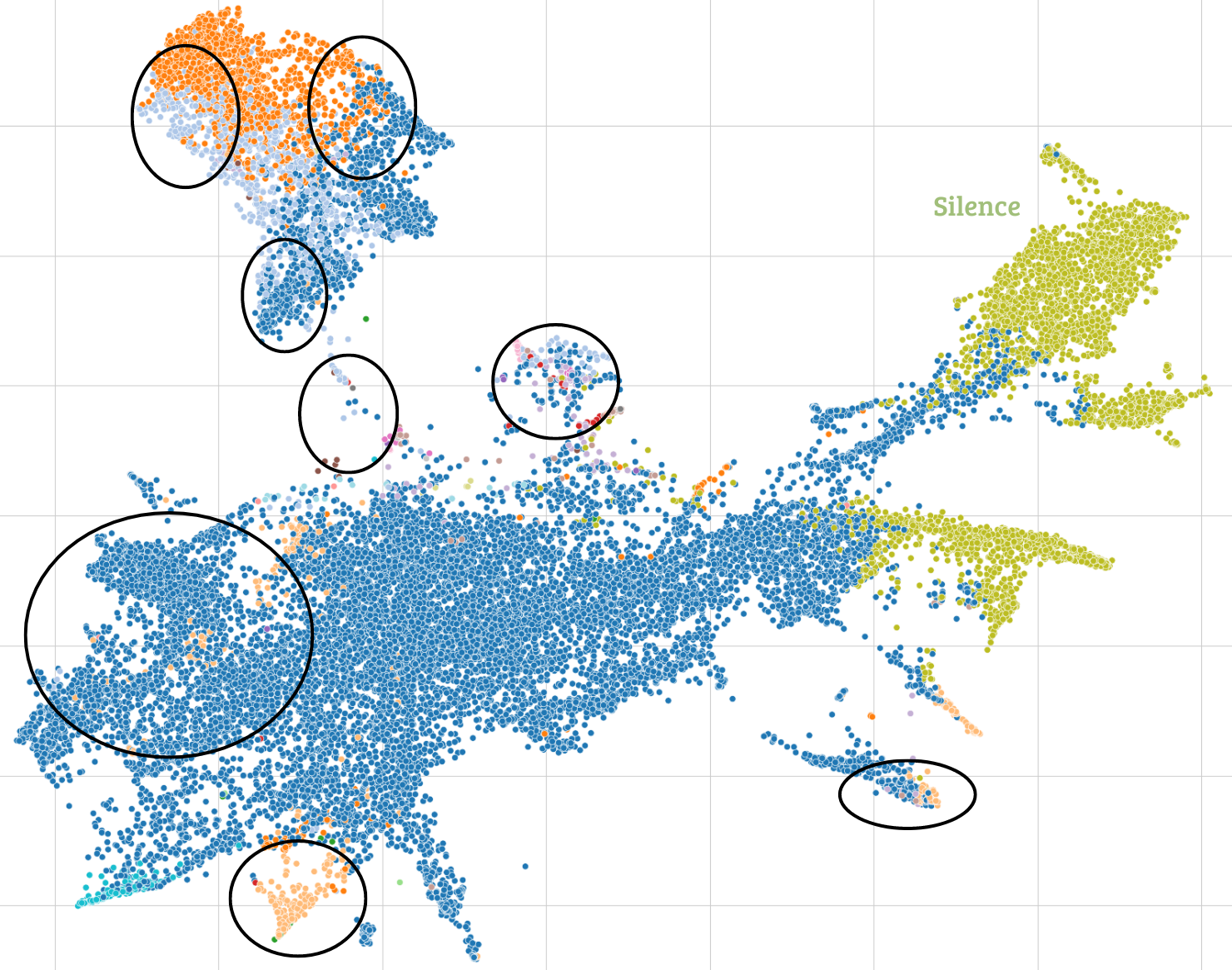}
        \caption{}
        \label{fig:subb_umap}
    \end{subfigure}
    \caption{(a) 2D UMAP representation of an AL iteration. The orange dots represent the medoids used in the iteration, the green stars correspond to the audios proposed for labeling and the remaining dots are the audios discarded for labeling. (b) 2D UMAP representation top1 pre-trained PANNs prediction. The different colors show distinct classes. The circles mark areas of interest according to the AL process.}
    \label{fig:umap}
\end{figure}

\begin{figure}
    \centering
    \includegraphics[width=1\linewidth]{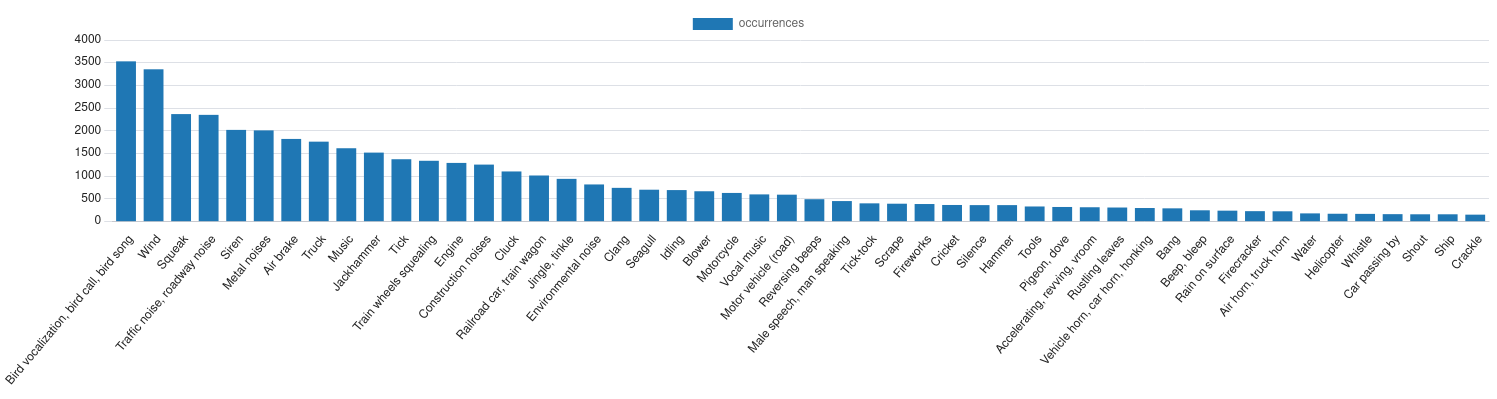}
    \caption{Histogram of tag frequencies for the 50 most repeated labels in the 6540 labeled audios}
    \label{fig:histogram}
\end{figure}

The labeling team was composed of 5 people divided into two groups of 3 and 2 members. In each AL iteration, 400 audios were proposed. Each group tagged 200 audios per week, 40 audios per day. Thus, the daily listening time corresponded to about 7 minutes per labeler, estimating a labeling effort of 30 minutes per day. Every Friday, a 30-minute meeting was held to unify criteria, propose suggestions and resolve doubts among taggers. The ontology used as a starting point was the Audioset ontology \cite{gemmeke2017audio}. If a labeler considered that an audio event could not be labeled with the available classes, a suggestion method was implemented so that the class became available in the next labeling iteration. These Friday meetings were also used to notify the rest of the labelers of the incorporation of new classes into the labeling process. In addition, a \textit{Doubt} class was created to be used in case a labeler was unsure which class to assign to a specific event. Every 10 labeling iterations, a doubt resolution iteration is performed. This iteration consisted in that each labeler had to re-listen and re-label their audios with some \textit{Doubt} event from the previous iterations and resolve them. The tagging process started on January 8, 2024. When this paper was drafted, 18 AL iterations and 1 doubt resolution iteration have been performed. The results obtained correspond to a total of 6540 strong labeled audios (due to vacations, not every week 400 audios could be labeled) which corresponds to a duration of 18 hours of labeled audio having employed 171 different labels (see Figure~\ref{fig:histogram} for insight about the tagged classes). The tagging effort corresponds to 52 hours per labeler. It should be noted that the creation of the groups can be parameterized and even (if deemed appropriate) different audios could be proposed to each labeler avoiding cross-validation, thereby increasing the number of labeled audios with the equivalent effort.

\begin{figure}
    \centering
    \includegraphics[width=0.9\linewidth]{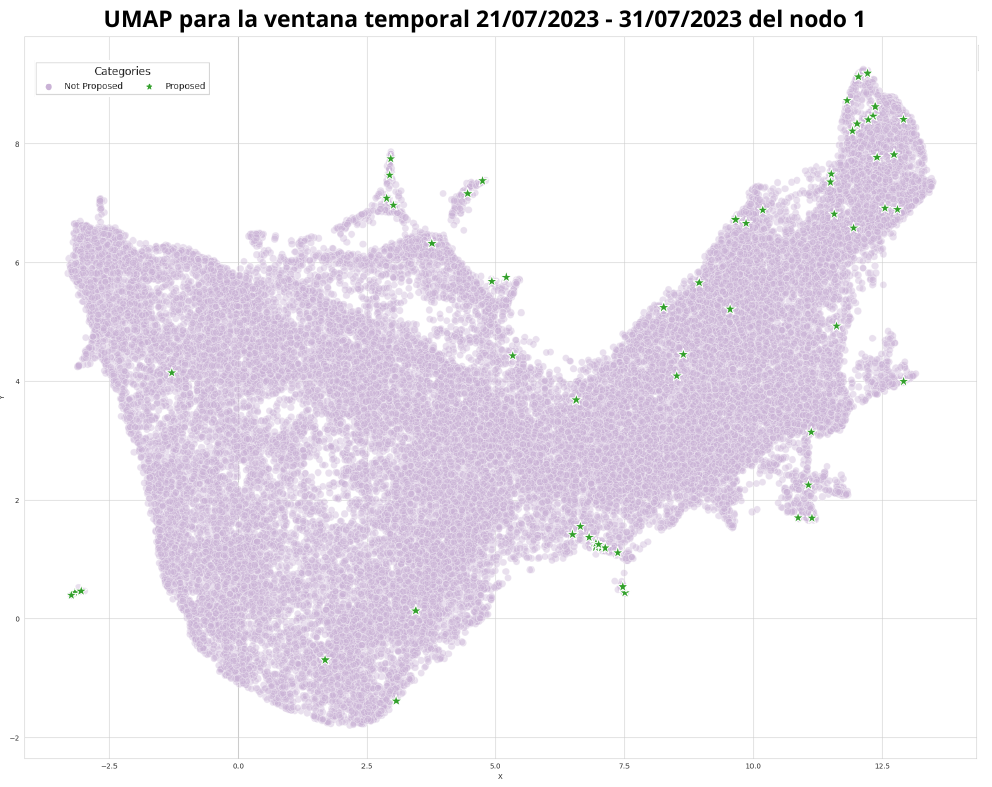}
    \caption{2D UMAP representation of the first AL iteration}
    \label{fig:first_al}
\end{figure}

Visual inspection using the UMAP dimensionality reduction algorithm has been used to validate the correct operation of the AL cycle. A concrete iteration of the AL process is shown in Figure~\ref{fig:suba_umap}. The medoids used in that iteration, the audios proposed for labeling and the discarded audios are plotted. In this particular iteration, the 10th, 400 audios have been proposed. As can be seen in the figure, the proposed audios are spread all over the feature space, some of them being close medoids. To analyze if this proposition makes sense, the same feature space is plotted in Figure~\ref{fig:subb_umap} but indicating the class provided by the CNN14 of PANNs pre-trained with Audioset. As can be appreciated (circled on the image) the AL solution proposes the border samples between two classes. In addition, audios that are classified as \textit{Silence} (green dots) have not been proposed in this iteration since medoids are available and they are not conflicting samples. By visually analyzing this iteration, it can be observed that the AL procedure is proposing relevant samples.

Figure~\ref{fig:first_al} shows the two-dimensional embeddings of the audios proposed by the AL algorithm in the first iteration. As can be seen in the image, no medoids are available and therefore the proposed audios are obtained by the MAL method \cite{8521336}.

\section{Conclusion}\label{sec:conclusions}

This paper establishes a complete framework for the creation of an audio database while optimizing the labeling budget in high-volume unlabeled data environments. This framework could be extremely useful when crowdsourcing is to be avoided to keep insight during the labeling process. The use case presented has been carried out in a commercial-industrial environment such as the port of Valencia (Spain). At the time of drafting this paper, 6540 10-second audios have been tagged with an effort of 52 hours per labeler having a pool of 5 labelers divided into two groups of 2 and 3 members.

As future work, continuing with the labeling process is expected to further increase the labeled database. Regarding the proposed Active Learning pipeline, the aim is to work on the algorithm itself, being able to propose new classifiers without label propagation because this technique could imply a problem once a large volume of labeled data is available. Finally, in the near future, it is intended to make public a subset of the labeled data for research use.

% A dataset is expected to be released soon as one of the outcomes of this work.

% \section{Purpose}
% The work proposes a framework to be used for Machine Listening projects in scenarios with constrained tagging budgets. General aspects of data collection, the AL-based labeling with AI-assisted techniques and the storage of the information obtained from the labeling process are presented.  In Machine Listening projects it is common that the largest and best-known sounding databases are either not suited to the particular problems of the project or cannot be used for commercial purposes. That is why a framework such as the one presented here should become standard practice in any AI-based machine listening project.

\section*{Acknowledgments and Disclosure of Funding}
The participation of Javier Naranjo-Alcazar, Jordi Grau-Haro and Pedro Zuccarello in this research was funded by the Valencian Institute for Business Competitiveness (IVACE) and the FEDER funds by means of project Soroll-IA2 (IMDEEA/2023/91). The research carried out for this publication has been partially funded by the project STARRING-NEURO (PID2022-137048OA-C44) funded by the Ministry of Science, Innovation and Universities of Spain and the European Union.



\begin{thebibliography}{6}
%


\bibitem{Schmid2023}
Schmid F., Morocutti T., Masoudian S., Koutini K., Widmer G.: Distilling the knowledge of transformers and CNNs with CP-mobile. In: Proceedings of the Detection and Classification of Acoustic Scenes and Events 2023 Workshop
(DCASE2023), pp. 161-165, (2023).

\bibitem{mesaros2021sound}
Mesaros A., Heittola T., Virtanen T., Plumbley M.D.: Sound event
detection: A tutorial. In: IEEE Signal Processing Magazine, vol. 38(5), pp. 67-83, (2021).

\bibitem{kandewal}
Khandelwal T., Kumar Das R., Siong Chng E., et al.: Sound event detection: A journey through dcase challenge series. In: APSIPA Transactions on Signal and Information Processing, vol. 13(1), (2024).

\bibitem{wilkinghoff2024self}
Wilkinghoff K.: Self-supervised learning for anomalous sound detection. In: ICASSP
2024-2024 IEEE International Conference on Acoustics, Speech and Signal Processing
(ICASSP), pp. 276-280. IEEE, (2024).

\bibitem{Aguado2023}
Aguado V., Navarro J., Vidaña-Vila E.: Learning in the wild: Bioacoustics
few shot learning without using a training set. In: Proceedings of the 8th Detection and
Classification of Acoustic Scenes and Events 2023 Workshop (DCASE2023), pp. 6-10,
Tampere, Finland, (2023).

\bibitem{mydlarz2019life}
Mydlarz C., Sharma M., Lockerman Y., Steers B., Silva C.,
Bello J.P.: The life of a new york city noise sensor network. In: Sensors, vol. 19(6):
pp. 1415, (2019).

\bibitem{Zinemanas2019}
Zinemanas P., Cancela P., Rocamora M.: Mavd: A dataset for sound event
detection in urban environments. In: Proceedings of the Detection and Classification of
Acoustic Scenes and Events 2019 Workshop (DCASE2019), pp. 263-267. New York
University, New York, (2019).

\bibitem{ooi2021strongly}
Ooi K., Watcharasupat K.N, Peksi S., Karnapi F.A., Ong Z.,
Chua D., Leow H., Kwok L., Ng X., Loh Z., et al.: A
strongly-labelled polyphonic dataset of urban sounds with spatiotemporal context. In:
2021 Asia-Pacific Signal and Information Processing Association Annual Summit and
Conference (APSIPA ASC), pp. 982-988. IEEE, (2021).

\bibitem{gemmeke2017audio}
Gemmeke J.F., Ellis D.PW., Freedman D., Jansen A., Lawrence W., R Channing Moore, Plakal M., Ritter M.: Audio set: An ontology and human-labeled
dataset for audio events. In: 2017 IEEE international conference on acoustics, speech and
signal processing (ICASSP), pp. 776-780. IEEE, New Orleans (2017).

\bibitem{piczak2015dataset}
Piczak K.J.: ESC: Dataset for Environmental Sound Classification. In: Proceedings of the 23rd Annual ACM Conference on Multimedia, pp. 1015-1018. ACM Press, (2015). 

\bibitem{salamon2014dataset}
Salamon J., Jacoby C., Bello J.P.: A dataset and taxonomy for urban sound research. In: Proceedings of the 22nd ACM international conference on Multimedia, pp. 1041-1044, (2014).

\bibitem{martin2023strong}
Martin-Morato I., Mesaros A.: Strong labeling of sound events using crowd-
sourced weak labels and annotator competence estimation. In: IEEE/ACM transactions on
audio, speech, and language processing, vol. 31, pp. 902-914, (2023).

\bibitem{lipping2019crowdsourcing}
Lipping S., Drossos K., Virtanen T.: Crowdsourcing a dataset of
audio captions. In: arXiv preprint, (2019) \url{https://arxiv.org/pdf/1907.09238}.

\bibitem{martin2021crowdsourcing}
Martin-Morato I., Harju M., Mesaros A.: Crowdsourcing strong labels
for sound event detection. In: 2021 IEEE Workshop on Applications of Signal Processing
to Audio and Acoustics (WASPAA), pp. 246-250. IEEE, New York, (2021).


\bibitem{settles2009active}
Burr Settles.: Active learning literature survey. (2009).

\bibitem{9217930}
Shuyang Z., Heittola T., Virtanen T.: Active learning for sound event detection. In: IEEE/ACM Transactions on Audio, Speech, and Language Processing, vol. 28, pp. 2895-2905, (2020). \url{https://doi.org/10.1109/TASLP.2020.302965}.

\bibitem{7952256}
Shuyang Z., Heittola T., Virtanen T.: Active learning for sound event
classification by clustering unlabeled data. In: 2017 IEEE International Conference
on Acoustics, Speech and Signal Processing (ICASSP), pp. 751-755, (2017). \url{doi: 10.1109/ICASSP.2017.7952256}.

\bibitem{8521336}
Shuyang Z., Heittola T., Virtanen T.: An active learning method using
clustering and committee-based sample selection for sound event classification. In: 2018
16th International Workshop on Acoustic Signal Enhancement (IWAENC), pp. 116-120, (2018). \url{doi:10.1109/IWAENC.2018.8521336.}


\bibitem{shishkin2021active}
Shishkin S., Hollosi D., Doclo S., Goetze S.: Active learning for sound
event classification using monte-carlo dropout and pann embeddings. In: Proceedings of
the 6th Workshop on Detection and Classication of Acoustic Scenes and Events (DCASE
2021), pp. 150-154. DCASE, (2021).

\bibitem{shishkin2024active}
Shishkin S., Hollosi D., Goetze S., Doclo S.: Active learning for sound
event classification using bayesian neural networks with gaussian variational posterior.
In: ICASSP 2024-2024 IEEE International Conference on Acoustics, Speech and Signal
Processing (ICASSP), pp. 896-900. IEEE, (2024).

\bibitem{Meire2023}
Meire M., Zegers J., Karsmakers P.: Active learning in sound-based
bearing fault detection. In: Proceedings of the 8th Detection and Classification of Acoustic Scenes and Events 2023 Workshop (DCASE2023), pp. 111-115. Tampere, Finland (2023).

\bibitem{8683063}
Wang Y., Mendez Mendez A. E., Cartwright M., Bello J.P.: Active
learning for efficient audio annotation and classification with a large amount of unlabeled data. In: ICASSP 2019-2019 IEEE International Conference on Acoustics, Speech and Signal Processing (ICASSP), pp. 880-884, (2019). \url{doi:10.1109/ICASSP.2019.8683063}

\bibitem{batch}
Citovsky G., DeSalvo G., Gentile C., Karydas L., Rajagopalan A., Rostamizadeh A., Kumar, S.: Batch active learning at scale. In: Advances in Neural Information Processing Systems, vol. 34, pp. 11933-11944, (2021).


\bibitem{kong2020panns}
Kong Q., Cao Y., Iqbal T., Wang Y., Wang W.,  Plumbley M.D.: Panns: Large-scale pretrained audio neural networks for audio pattern recognition.
In: IEEE/ACM Transactions on Audio, Speech, and Language Processing, vol. 28, pp. 2880-2894, (2020).












% \bibitem {smit:wat}
% Smith, T.F., Waterman, M.S.: Identification of common molecular subsequences.
% J. Mol. Biol. 147, 195?197 (1981). \url{doi:10.1016/0022-2836(81)90087-5}

% \bibitem {may:ehr:stein}
% May, P., Ehrlich, H.-C., Steinke, T.: ZIB structure prediction pipeline:
% composing a complex biological workflow through web services.
% In: Nagel, W.E., Walter, W.V., Lehner, W. (eds.) Euro-Par 2006.
% LNCS, vol. 4128, pp. 1148?1158. Springer, Heidelberg (2006).
% \url{doi:10.1007/11823285_121}

% \bibitem {fost:kes}
% Foster, I., Kesselman, C.: The Grid: Blueprint for a New Computing Infrastructure.
% Morgan Kaufmann, San Francisco (1999)

% \bibitem {czaj:fitz}
% Czajkowski, K., Fitzgerald, S., Foster, I., Kesselman, C.: Grid information services
% for distributed resource sharing. In: 10th IEEE International Symposium
% on High Performance Distributed Computing, pp. 181?184. IEEE Press, New York (2001).
% \url{doi: 10.1109/HPDC.2001.945188}

% \bibitem {fo:kes:nic:tue}
% Foster, I., Kesselman, C., Nick, J., Tuecke, S.: The physiology of the grid: an open grid services architecture for distributed systems integration. Technical report, Global Grid
% Forum (2002)

% \bibitem {onlyurl}
% National Center for Biotechnology Information. \url{http://www.ncbi.nlm.nih.gov}


\end{thebibliography}
\end{document}